\documentclass[sn-mathphys-num]{sn-jnl}

\usepackage{multirow}%
\usepackage{amsmath,amssymb,amsfonts}%
\usepackage{amsthm}%
\usepackage{mathrsfs}%
\usepackage[title]{appendix}%
\usepackage{xcolor}%
\usepackage{textcomp}%
\usepackage{manyfoot}%
\usepackage{booktabs}%
\usepackage{algorithm}%
\usepackage{algorithmicx}%
\usepackage{algpseudocode}%
\usepackage{listings}%

\usepackage[varg]{txfonts}   
\usepackage{graphics}
\usepackage[utf8x]{inputenc}
\usepackage{ulem}

\usepackage{xcolor}

\usepackage{todonotes}

\usetikzlibrary{decorations.pathmorphing}
\def\hhresp{
\begin{tikzpicture}[thick]
\draw[thick] (0.2,0) - - (1.65,0);
\draw[thick] (1.45,-0.15) - - (1.45,0.15);
\end{tikzpicture}
}
\def\hh{
\begin{tikzpicture}[thick]
\draw[thick] (0.2,0) - - (1.65,0);
\draw[white] (0.2,-0.15) - - (1.65,-0.15);
\end{tikzpicture}
}
\def\vv{
\begin{tikzpicture}[thick]
\tikzset{snake it/.style={decorate, decoration=snake}}
\path [draw=black,snake it] (0,0) - - (1.45,0);
\end{tikzpicture}
}
\def\vvresp{
\begin{tikzpicture}[thick]
\tikzset{snake it/.style={decorate, decoration=snake}}
\path [draw=black,snake it] (0,0) - - (1.45,0);
\draw[thick] (1.285,-0.15) - - (1.285,0.15);
\end{tikzpicture}
}

\begin{document}

\title[Field Theoretic Renormalization Group in a Model of Random Surface Growth]{Field Theoretic Renormalization Group in an Infinite-Dimensional Model of Random Surface Growth in Random Environment}


\author[1,2]{\fnm{N~V~} \sur{Antonov}}\email{n.antonov@spbu.ru}
\equalcont{These authors contributed equally to this work.}

\author[1]{\fnm{A~A~} \sur{Babakin}}\email{st068365@student.spbu.ru}
\equalcont{These authors contributed equally to this work.}

\author[1,2]{\fnm{N~M~} \sur{Gulitskiy}}\email{n.gulitskiy@spbu.ru}
\equalcont{These authors contributed equally to this work.}

\author*[1]{\fnm{P~I~} \sur{Kakin}}\email{p.kakin@spbu.ru}
\equalcont{These authors contributed equally to this work.}

\affil*[1]{\orgdiv{Department of Physics}, \orgname{Saint Petersburg State University}, \orgaddress{\street{7/9 Universitetskaya nab.}, \city{Saint~Petersburg}, \postcode{199034}, \country{Russia}}}

\affil[2]{\orgdiv{N.N. Bogoliubov Laboratory of Theoretical Physics}, \orgname{Joint Institute for Nuclear Research}, \orgaddress{\city{Dubna}, \postcode{141980}, \state{Moscow Region}, \country{Russia}}}


\abstract{The influence of a random environment on the dynamics of a fluctuating rough surface is investigated using a field theoretic renormalization group. The environment motion is modelled by the stochastic Navier--Stokes equation, which includes  
both a fluid in thermal equilibrium and a turbulent fluid. The surface is described by the generalized Pavlik's stochastic equation. As a result of fulfilling the renormalizability requirement,
the model necessarily involves an infinite number of coupling constants. The one-loop counterterm is derived in an explicit closed form. The corresponding renormalization group equations demonstrate the existence of three two-dimensional surfaces of fixed points in the infinite-dimensional parameter space.
If the surfaces contain IR attractive regions, the problem allows for the large-scale, long-time scaling behaviour.
For the first surface (advection is irrelevant) the critical dimensions of the height field $\Delta_{h}$, the response field  $\Delta_{h'}$ and the frequency $\Delta_{\omega}$ are non-universal through the dependence on the effective couplings. For the other two surfaces (advection is relevant) the dimensions are universal and they are found exactly.}




\maketitle

\section{Introduction}\label{sec:Intro}

Kinetic roughening of randomly growing interfaces is an important example of dynamic 
non-equilibrium behaviour both conceptually (because it reveals self-similar long-range spatio-temporal correlations without any external tuning) and in relation to possible applications. 

The wide range of physical systems with random growth includes chemical solutions where substance deposition creates phase boundary, propagating flames, smoke or solidification fronts, vicinal surfaces, populations of cells, crystals undergoing molecular beam epitaxy, and many others; see, e.g. \cite{Eden,Yan2,Yan,Zhang,Stanley,Barabasi,Xia} and references therein.

As a rule, various correlation functions for those processes exhibit scaling (self-similar or self-affine) behaviour in the infrared range (IR, long times, large distances), for example: 
\begin{equation}
C_n (t,r) = \langle \left[ h(t,{\bf x})- h(0,{\bf 0}) \right]^n \rangle
\simeq r^{n\chi} F_n(t/r^z), \quad r=|{\bf x}|.
\label{scaling}    
\end{equation}
Here $h=h(t,{\bf x})$ is a relevant field (e.g. a fluctuating part of the surface height), the brackets denote averaging over the statistical ensemble,
$\chi$ and $z$ are the widely used notations for the pair of critical exponents, and $F_n(\cdot)$ are scaling functions.

One of theoretical approaches to the study of kinetic roughening consists of establishing scaling relations like (\ref{scaling}) on the base of certain (usually semi-phenomenological) dynamical models, calculating the critical exponents in a systematic way and investigating their universality (that is, dependence on the dimension of space $d$ and other parameters of the model). 

One such model is the celebrated Kardar--Parisi--Zhang (KPZ) equation \cite{KPZ} that can arguably be considered ``the Ising model for non-equilibrium phenomena'':

\begin{equation}
\partial_t h = \kappa_0 \partial^2 h + U(h) +\eta .   
\label{equation}    
\end{equation}
Here $h=h(x)$ is the fluctuating part of the height field,
$\kappa_0>0$ is the surface tension coefficient, $U(h)$ is a certain non-linearity and $\eta=\eta(x)$ is the random noise. Here and below, 
$x=\{t,{\bf x}\}$, ${\bf x}=\{x_1,\dots,x_d\}$,  
$d$ is the dimension of space, 
$\partial_t=\partial/\partial t$,
$\partial_i=\partial/\partial x_i$,
$\partial^2=\partial_i\partial_i$ is the Laplace operator;
summation over repeated vector indices is always implied. The noise is Gaussian with the correlation 
function\footnote{The noise $\eta$ is supposed to have a certain constant component $\langle \eta(x)\rangle$
that guarantees that $\langle h(x)\rangle=0$, which follows from the meaning of $h$ as a fluctuating part, but in practical calculations they can simply be simultaneously ignored.}
\begin{equation}
\langle \eta(x)\eta(x')\rangle_{\eta} = B_0\, \delta(x-x') =
B_0\, \delta(t-t')\, \delta({\bf x}-{\bf x}'), \quad B_0>0.
\label{noise}    
\end{equation}
In the KPZ model, the non-linearity is taken in the form\footnote{To be precise, this model was introduced earlier in \cite{FNS1} by Forster, Nelson and Stephen in terms of a potential vector field $v_i = \partial_i h$ as a stochastic $d$-dimensional generalization of the Burgers equation.}
\begin{equation}
U(h)= \frac{\lambda_0}{2}\, (\partial h)^2 =
\frac{\lambda_0}{2}\,(\partial_i h)\,(\partial_i h),
\label{KPZ}    
\end{equation}
where ${\lambda_0}$ can be of either sign and models lateral growth or erosion.

One of the most powerful tools of studying scaling behaviour is the renormalization group~(RG) analysis; see, e.g. the monograph
\cite{Vasiliev} and literature cited there. The RG was applied to the stochastic problem (\ref{equation})~-- (\ref{KPZ}) in the very first papers \cite{KPZ,FNS1} in the form of Wilson's recursion relations. Later, the more advanced field theoretic RG, suitable for higher-order calculations, was employed; see \cite{Wiese2,Lassig,Wiese} for the references and the discussion.

A large number of models based on the original KPZ equation but with various adjustments and modifications have been proposed: 
``colored'' noise $\eta$ with finite correlation time
\cite{Modify1,Modify2}, 
``quenched'' $h$-dependent and time-independent noise \cite{Modify7,Korea},
vector or matrix field $h$ \cite{Modify4,Modify5,Modify6},
random coupling constant \cite{Modify9},
inclusion of superdiffusion \cite{Modify3},
inclusion of anisotropy \cite{Modify7,Wolf,Modify8},
conserved field $h$ \cite{Modify10,Modify11,Modify12},
modified non-linearity  $U(h)$ \cite{Pavlik} and so on.

The last one on that list was introduced by Pavlik \cite{Pavlik} and features non-linear term in~(\ref{equation}) of the form 
$U(h) \simeq \partial^2 h^2/2 = (\partial h)^2 + h \partial^2 h$. 
The first term is the original KPZ non-linearity~(\ref{KPZ}), while the second one can be viewed as an $h$-dependent contribution to the surface tension, which makes it essentially non-linear.

However, the Pavlik's model with the single non-linearity $U(h) \simeq \partial^2 h^2/2$ is not self-sufficient in the following sense. The dimensional analysis along with more sophisticated renormalization arguments show that an infinite number of non-linear terms $\partial^2 h^n$ with $n\ge2$ must be included into the model because they all are equally relevant~\cite{AV}. Thus, the model should be extended to involve infinitely many coupling constants.

It is necessary to note here that the model nearly identical to the extended Pavlik's model was proposed earlier and independently by Diaz-Guilera 
 in an attempt to construct a continuous model of 
 self-organized criticality \cite{Diaz-Guilera}.
 Later on, its modifications where the random noise is described by a Gaussian ensemble with finite correlation time was proposed and studied in \cite{Volk1,Volk2}.

The Pavlik's model can also be viewed as an isotropic version of Pastor-Satorras--Rothman model of landscape erosion \cite{Pastor1,Pastor2} that requires a similar infinite extension \cite{AK1,AK2,Static,Duclut}. It also has some similarity with the anisotropic Hwa-Kardar equation \cite{SOC1,SOC2} where the non-linearity has the form of a total derivative. This means that both equations turn into continuity equations in the absence of noise, just like the aforementioned models of landscape erosion \cite{Pastor1,Pastor2}.

Although Pavlik's model and its extended relatives were originally proposed for kinetic roughening \cite{Pavlik,AV} and self-organized criticality \cite{Diaz-Guilera}, they also describe strongly non-linear diffusion; see, e.g \cite{JETP,ABK,Symm}. In this regard, it is worth noting that the need to include arbitrarily high powers of the scalar field in the diffusion equation was suggested as early as 1937~\cite{Pourous}.

The extended versions of Pavlik's model with infinitely many couplings have recently been used to describe critical activity of brain neurons in~\cite{Stapmanns,Tiberi}.

It is well-known that critical behaviour of equilibrium systems is
highly sensitive to various external perturbations such as, for example, environment motion. The same is true for non-equilibrium dynamical systems where laminar or turbulent flow can ``level'' scaling behaviour making it trivial (the mean-field one) or, on the contrary, give rise to new regimes of it~\cite{Onuki,Nelson,Satten,Nandy,AHH}. What is more, such motion is difficult to exclude in experimental settings. That is why it is important to account for  its effects when considering models of kinetic roughening and non-linear diffusion. For the KPZ model, this problem was explored in the series of works \cite{AK3,AKL,AGKK,Reiter}.

In this paper, we analyse the extended Pavlik's model within a field-theoretic RG analysis taking into account random motion of the environment. The latter is described by a stochastic differential Navier--Stokes (NS) equation for an incompressible viscous fluid with a special choice of the external random force that allows one to consider both a medium in thermal equilibrium and a strongly turbulent fluid.
The original formulation of the model involves two coupled stochastic differential equations.
It can be reformulated as a certain field-theoretic model that includes an infinite number of coupling constants.

Description of the model and construction of the action functional are given in Section~2.
In Section~3, UV divergences of the model are analysed. It is shown that the model can be considered as multiplicatively renormalizable in the infinite space of coupling constants.
In the central  Section~4, the one-loop counterterm is derived in an explicit closed form. This allows us to derive the RG equations and to give the explicit expressions for their  coefficients: the anomalous dimensions $\gamma$ and the $\beta$ functions; see  Section~5.

In Section~6, we  establish that the RG equations have three two-dimensional surfaces of fixed points in the infinite-dimensional parameter space (in contrast with conventional field-theoretic models with 
their single fixed points). If these surfaces involve IR attractive regions, the model 
demonstrates IR (large-scale, long-time) asymptotic scaling behaviour.

The first surface corresponds to the situation where the influence of the environment is
irrelevant; then the scaling dimensions are non-universal in the sense that they depend on the specific point on the surface.

For the other two surfaces (that correspond to the fluid in equilibrium and the turbulent fluid) the dimensions are shown to be universal. Moreover, they are given exactly by the leading (one-loop) approximation of the corresponding $\varepsilon$ expansion.
Conclusion and discussion are given in Section~7.

\section{Description of the model and its field theoretic formulation} \label{sec:Model}

Extended Pavlik's model \cite{AV,Diaz-Guilera} of a fluctuating surface is described by the
equations (\ref{equation}), (\ref{noise})
where the former is taken in the form
\begin{equation}
\partial_t h = \partial^{2} V(h) + \eta,
\label{advection}
\end{equation}
with function $V(h)$ defined as an infinite series in powers of $h$:
\begin{equation}
V(h) = \sum_{n=1}^{\infty}\, \frac{1}{n!}\, \lambda_{n0}\, h^n;
\label{V}
\end{equation}
in the notation of equation (\ref{equation}), $\kappa_0=\lambda_{10}$.

The advection by turbulent environment is introduced by the ``minimal'' replacement \mbox{$\partial_t \to \nabla_t$} in the equation~(\ref{advection}), where
\begin{equation}
\nabla_t=\partial_t + ({v} {\partial})=\partial_t + {v}_k \partial_k,
\label{nabla}
\end{equation}
is the Lagrangian (Galilean covariant) derivative.

The turbulent medium is modelled by the stochastic NS equation for an incompressible viscous fluid:
\begin{equation}
{\nabla}_{t}{{v}}_{i} - {\nu}_{0}{\partial}^{2}{{v}}_{i} + {\partial}_{i}\wp - {f}_{i} = 0.
\label{NSeq}
\end{equation}
Here ${\bf v}=\{v_i\} $ is the transverse (due to the incompressibility condition $\partial_i{{v}}_{i} = 0$) velocity field,  $\wp$ is the pressure, ${\bf f}=\{f_i\}$ is the transverse random force per unit mass (all these quantities depend on $x = \left(t, {\bf x}\right)$) and $\nu_{0}$ is the kinematic coefficient of viscosity. 

The force ${\bf f}$ is assumed to be Gaussian with zero mean and a given pair correlation function:
\begin{equation}
\langle f_i(x) f_j(x') \rangle_{f} = \delta(t-t') \int_{k>m} 
\frac{d{\bf k}}{(2\pi)^d}\, P_{ij}({\bf k}) D_f(k) \, \exp \{{\rm i} {\bf k}({\bf x-x'})\},
\label{noiseNS}
\end{equation}
where $P_{ij}({\bf k}) = \delta_{ij} - k_i k_j / k^2$ is the transverse 
projector, $k\equiv |{\bf k}|$ is the wave number and 
\begin{equation}
D_f(k) = D_{0}k^{4 - d - y} + D'_{0} k^2, \quad D_{0}>0,\, \ D'_{0} >0.
\label{noiseNS1}
\end{equation}

Let us list the reasons behind this choice of the correlation function. The power-like form~(\ref{noiseNS1}) with $D'_{0}=0$ is typical for the standard field theoretic approach to the fully developed turbulence; see, e.g. the monographs~\cite{Vasiliev,Red}, the review paper~\cite{UFN} and 
references therein. The physical value of the exponent $y$ corresponds to the limit $y\to4$, where the function~(\ref{noiseNS1}) with $D'_{0}=0$ and an appropriate choice  of the amplitude $D_{0}$ can be viewed as a power-like representation of the function $\delta({\bf k})$ that describes the energy pumping by a large-scale stirring. The model becomes logarithmic at $y = 0$ so that the exponent $y$ plays the role of the formal expansion parameter in the RG perturbation theory.

However, the model (\ref{advection}), (\ref{V})  (as well as the KPZ model) becomes logarithmic at $d=2$~\cite{AV}, 
and its RG analysis should be performed within the expansion in $\varepsilon=2-d$. In order to make the RG analysis of the full model internally consistent, it is necessary to treat $y$ and $\varepsilon$ as small parameters of the same order.

In its turn, the RG analysis of the model (\ref{NSeq})~-- (\ref{noiseNS1})
near $d=2$ becomes rather delicate~\cite{Red,HoNa,AntiRonis}. 
It shows that, in order to ensure renormalizability, a local term must be added into the correlation function of the random force, namely $D'_{0} k^2$ in~(\ref{noiseNS1}). It is this term that bears the renormalization constant,
while the original non-local term with the amplitude $D_{0}$ remains intact.

At $D_{0}=0$, the model (\ref{NSeq})~-- (\ref{noiseNS1}) becomes local and
describes a fluid in thermal equilibrium; it was introduced in \cite{FNS1,FNS} in connection to the problem of ``long tails'' in derivation of hydrodynamics equations. It is renormalizable in itself and logarithmic at $d=2$.

Thus, the general model (\ref{noiseNS1}) involves a turbulent fluid as well as a fluid in equilibrium.

The ultraviolet (UV) divergences take on the form of singularities in $y$, $\varepsilon$ and their combinations, while the coordinates of the fixed points and various critical dimensions are calculated as expansions in $y$ and $\varepsilon$ with the assumption that $y\sim \varepsilon$.
\footnote{They are not simple double series in two parameters: the expressions like $y^2/(\varepsilon-y)$ can also appear (see, e.g. equation (\ref{IRGS3}) in Section~\ref{sec:scaling} below) and should be treated as quantities of order $\varepsilon\sim y$. See \cite{AntiRonis} for a more detailed discussion.}

According to the general De~Dominicis–Janssen theorem (see, e.g. Chapter 5 in \cite{Vasiliev} and references therein), the original stochastic problem (\ref{advection})~-- (\ref{noiseNS1}) can be represented as the field theoretic model of an extended set of fields $\Phi = \left\{h, h', {v}, {v}'\right\}$ with the action functional ${\cal S}(\Phi) = {\cal S}_{h}(\Phi)+{\cal S }_{{v}}(\Phi)$ where
\begin{equation}
{\cal S}_{h}(\Phi)	= \frac{1}{2} h' h' +
h' \left[ - \nabla_t h + \partial^2 V(h)
\right],
\label{Action}
\end{equation}
\begin{equation}
	{\cal S}_{{v}}(\Phi) = \frac{1}{2} {{v}}' D_{f}  {{v}}' +
 {{v}}' \left[ - \nabla_t  {{v}} + {\nu}_{0}\partial^2 {{v}}
\right].
\label{NSAction}
	\end{equation} 
The function $D_f$ is the one in expressions (\ref{noiseNS}), (\ref{noiseNS1}).

Here and below, all the needed integrations over the arguments $x=\{t, {\bf{x}}\}$ and summations over repeated vector indices are implied for all terms such as the ones in equations (\ref{Action}), (\ref{NSAction}). For example,
\begin{equation}
h' \nabla_{t} h = \int{dt \, \int{d{\bf{x}} h'(t, {\bf{x}}) \nabla_{t} h(t,{\bf{x}})}}.
    \label{example}
\end{equation}

The field theoretic formulation allows one to represent various correlation and response functions of the original stochastic problem as functional averages over the full set of fields $\Phi = \{h, h',{v}, {v'}\}$ with the weight $\exp{\cal S}$ (for more details, see \cite{Vasiliev}). 

Now the field theoretic tools that include Feynman diagrammatic techniques, functional equations of Schwinger–Dyson type, Ward identities for Galilean symmetry, renormalization theory and RG can be applied to the study of the field theoretic model (\ref{Action}), (\ref{NSAction}).

\begin{figure}[h!]
\usetikzlibrary{decorations.pathmorphing}
\tikzset{snake it/.style={decorate, decoration=snake}}
\minipage{0.3333\textwidth}
\begin{center}
\begin{tikzpicture}[thick]
\draw[thick] (0.2,0) - - (1.45,0);
\draw[thick] (1.25,-0.15) - - (1.25,0.15);
\fill[black] (1.45,0) circle(2pt);
\draw[thick] (1.45,0) - - (2.6,0.65);
\path[draw=black,snake it] (1.45,0) - - (2.6,-0.65);
\end{tikzpicture}
\end{center}
\endminipage
\hfill
\minipage{0.3333\textwidth}
\usetikzlibrary{decorations.pathmorphing}
\begin{center}
\begin{tikzpicture}[thick]
\draw[thick] (0.2,0) - - (1.45,0);
\draw[thick] (1.25,-0.15) - - (1.25,0.15);
\fill[black] (1.45,0) circle(2pt);
\draw[thick] (1.45,0) - - (2.55,0.55);
\draw[thick] (1.45,0) - - (2.55,-0.55);
\draw[thick] (1.45,0) - - (2.65,0.3);
\draw[dashed,thick] (1.45,0) - - (2.65,-0.3);
\draw[thick] (1.45,0) - - (2.7,0);
\end{tikzpicture}
\end{center}
\endminipage
\hfill
\minipage{0.3333\textwidth}
\usetikzlibrary{decorations.pathmorphing}
\begin{center}
\begin{tikzpicture}[thick]
\path [draw=black,snake it] (0.2,0) - - (1.45,0);
\draw[thick] (1.25,-0.15) - - (1.25,0.15);
\fill[black] (1.45,0) circle(2pt);
\path [draw=black,snake it] (1.45,0) - - (2.6,0.65);
\path [draw=black,snake it] (1.45,0) - - (2.6,-0.65);
\end{tikzpicture}
\end{center}
\endminipage
\caption{From left to right: the vertex $h'(v\partial)h$, vertices $h'\partial^2 V(h)$ (the dotted line stands for the remaining $(n-4)$ tails without strokes), the vertex $v'(v\partial)v$.}
\label{FeynmanVert}
\end{figure}
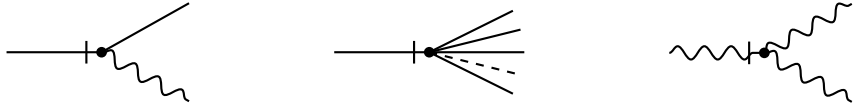

In the diagrammatic technique, the propagators for that model will be denoted by the following lines:
\begin{eqnarray}
\langle hh' \rangle_{0} = \hhresp \, , & \quad \langle hh \rangle_{0} = \hh \, ,
    \label{D1}
\end{eqnarray}
\begin{eqnarray}
\langle v_i v'_j \rangle_{0} = \vvresp \, , & \quad \langle v_i v_j \rangle_{0} = \vv \,.
    \label{D2}
\end{eqnarray}
The vertices are shown in the figure~\ref{FeynmanVert}: 
the vertices entering $h'\partial^2 V(h)$ have one tail with a stroke and a number of tails without it. The vertex $h'(v\partial)h$ has one wavy tail, a tail with a stroke and a tail without it. The vertex $v'(v\partial)v$ features only wavy tails.

\section{UV divergences and renormalization} \label{sec:UV}

\subsection{Canonical dimensions} 
\label{sec:Canonical}

The Green's functions for the model (\ref{Action}), (\ref{NSAction}) involve UV divergences, which can be eliminated by the standard renormalization procedure. 

The presence of UV divergences can be established through the canonical dimensions analysis of fields and parameters; see, e.g. Sections $1.15$, $1.16$ in \cite{Vasiliev}. In contrast to static models that essentially include only one dimensional scale, dynamic ones contain two independent scales: spatial scale $[L]$ and temporal scale $[T]$ (see Sections $1.17$ and $5.14$ in \cite{Vasiliev}). This means that the canonical dimension of any quantity $F$ are specified with two numbers: the momentum dimension $d_F^k$ and the frequency dimension ${d_F^{\omega}}$,
\begin{equation} 
\left[F\right] \sim \left[T\right]^{-d_F^\omega} \left[L\right]^{-d_F^k}.
\nonumber
\end{equation}
Eventually, the dimensions of all quantities are found from the standard normalization conditions:
\begin{equation}
	d_{\bf k}^k = -d_{\bf x}^k = 1, \quad d_{\bf k}^\omega = d_{\bf x}^\omega = 0, \quad d_\omega^k = d_t^k = 0, \quad d_{\omega}^{\omega}=-d_t^{\omega}=1,
	\nonumber
\end{equation}
and also from the requirement that all the terms in the action functional be dimensionless with respect to each scale. 

Introducing the total canonical dimension $d_F = d_F^k + 2d_F^{\omega}$ is also necessary since it subsequently plays the same role as the conventional momentum dimension does in static models. It is chosen so that the viscosity coefficients $\nu_0$ and parameter $\lambda_{10}$ appear dimensionless and $\partial_t \sim \partial^2$, where $\partial_t$, $\partial^2$ are derivatives entering the free parts of the action functionals (\ref{Action}), (\ref{NSAction}); hence the coefficient $2$ in $d_F$; see Section $5.14$ in \cite{Vasiliev}. 

Since the dimensions of all quantities in the action functionals must be determined unambiguously, one should get rid of unnecessary parameters, if any. In equation (\ref{advection}), the amplitude $B_0$ from expression (\ref{noise}) can be eliminated by an appropriate rescaling of the fields and the parameters.
In other words, one can set $B_0 = 1$ with no loss of generality, which is assumed in what follows. Such rescaling naturally does not affect critical exponents in expression (\ref{scaling}).

Canonical dimensions of the other parameters and the fields are given in the Table~\ref{t1}. 
Dimensions of their renormalized counterparts (without the subscript "0") and the renormalization mass $\mu$ (additional parameter of the renormalized theory) are also included there.  

\begin{table}[h!]
\caption{Canonical dimensions of the fields and the parameters in the model (\ref{Action}),  (\ref{NSAction}).}
\label{t1}
\begin{tabular}{c||c|c|c|c|c|c|c|c}
     \hline\hline
     $F$ & $h$   &  $h'$  & ${v}$  & ${v}'$ &  $m$, $\mu$  & ${D}_{0}$ & $D'_{0}$ &  ${\lambda}_{n0}$  \\  \hline 
 $d^{k}_{F}$ &  $1 - \varepsilon/2$ &  $1 - \varepsilon/2$      &     $ - 1$   &   $3 - \varepsilon $  &   $1$              & $y - 6$ & $\varepsilon - 6$ & $\left(1 - \varepsilon/2\right)\left(1 - n\right) - 2$  \\  
$d^{\omega}_{F}$ &  $- 1/2$ &  $1/2$      &     $1$ & $-1$ &  $0$    & $3$ & $3$ & $\left(n + 1\right)/2$  \\  
$d_{F}$ &  $- \varepsilon/2$ &  $2 - \varepsilon/2$   &  $1$ & $ 1 - \varepsilon$  &   $1$   &  $y$ & $\varepsilon$ & $\varepsilon\left(n - 1\right)/2$   \\  
 \hline\hline
$F$          & ${\lambda}_{n}$       & $g_{0}^{\prime}$  &    ${\nu}_{0}$, ${\nu}$  & $g_{0}$ & $g_{n0}$ & $g$ & $g^{\prime}$ & $g_{n}$ \\  \hline
 $d^{k}_{F}$ &   $ - \left(n + 1\right)$ & $\varepsilon$  & $-2$                          & $y$         & $\varepsilon\left(n - 1\right)/2$         & $0$ & $0$ & $0$  \\  
$d^{\omega}_{F}$  &  $\left(n + 1\right)/2$ & $0$ &  $1$    & $0$         & $0$         & $0$  & $0$ & $0$ \\  
$d_{F}$ & $0$ & $\varepsilon$ & $0$ & $y$ & $\varepsilon\left(n - 1\right)/2$ & $0$ & $0$ & $0$ \\ \hline\hline
\end{tabular}
\end{table}

We also introduced charges $g_{n0}$, $g_{0}$ and $g'_{0}$ and their renormalized counterparts: 
\begin{eqnarray}
    \lambda_{n0} &=& g_{n0} \nu_{0}^{(n+1)/2}, \quad
\lambda_{n} = g_{n} {\nu}^{(n+1)/2} {\mu}^{\varepsilon (n-1)/2} \quad (n > 0), \\
D_{0} &=& g_{0}\nu_{0}^{3}, \quad D = g\nu^3\mu^y \\ 
D'_{0} &=& g'_{0}\nu^{3}_{0}, \quad D' = g'\nu^3\mu^{\varepsilon}.
    \label{charges1}
\end{eqnarray}

The charges are chosen so that $d_{g_{n0}}^{\omega} = d_{g_{0}}^{\omega} = d_{g'_{0}}^{\omega} = 0$, while the corresponding renormalized counterparts are completely dimensionless. The charges $g_0$, $g'_0$ and $g_{n0}$, $n>1$, serve as expansion parameters in the ordinary perturbation theory for the original unrenormalized model. 
We are interested in the expansion in the number of loops, then the relations $g_{n0} \sim g_{20}^{(n-1)}$ should be implied for internal consistency. The dimensionless ratio $g_{10} \sim~1$ of two kinematic coefficients is not an expansion parameter but must be considered alongside the other charges.

From Table~\ref{t1}, it follows that all interaction vertices become logarithmic when $y=0$ and $\varepsilon=0$ as the corresponding charges $g_0$, $g'_0$ and $g_{n0}$ become dimensionless. That means that the exponents $y$ and $\varepsilon$ measure deviation from logarithmicity and according to the general ideology of renormalization they should be considered as formal small parameters of the same order $(y \sim \varepsilon)$. 

\subsection{UV divergences and counterterms} 
\label{sec:UVD}
Superficial UV divergences (which removal requires counterterms) are present in the 1-irreducible Green's functions for which the formal index of divergence
\begin{equation}
\delta = (d+2) - \sum_{\Phi}\, N_{\Phi}\, d_{\Phi}
\label{index}
\end{equation}
is a non-negative integer in the logarithmic theory ($y=\varepsilon=0$). Here $N_{\Phi}$ are the numbers of the fields
$\Phi=\{h', h, v', v\}$ entering the Green's function and $d_{\Phi}$ are their total canonical dimensions; see, e.g. \cite{Vasiliev} (sec.~5.15), \cite{Red} (sec.~1.4) and \cite{UFN}.

Additionally, when analysing divergences in the model (\ref{Action}), the following considerations should be taken into account as well: 

${\bf (1)}$ For any dynamical model of type (\ref{Action}) and (\ref{NSAction}), all 1-irreducible functions that do not include the response fields
${v}'$ and $h'$,  contain closed loops of retarded propagators and vanish.
Thus, it suffices to only consider functions with $N_{v'}+N_{h'} >0$. 

${\bf (2)}$ At the vertex ${v}'({v}\partial){v}$, the derivative $\partial$ can be carried over by integration by parts to the field ${v}'$ (due to the incompressibility condition). Thus, in any 1-irreducible diagram, for each external field ${v}'$ attached to this vertex, the corresponding external momentum is allocated, and the real divergence index of the diagram decreases by the corresponding number of units: $\delta'=\delta - N_{{v}'}$. At the same time, the field ${v}'$ is required to enter the corresponding counterterms only under spatial derivative. The same is true for the vertex $h'({v} \partial)h$: $\delta'=\delta - N_{h'}$.

${\bf (3)}$ In the vertices $h'\partial^2 h^n$, the Laplace operator can also be moved onto the field $h'$ with integration by parts. Thus, 
$\delta'=\delta - 2N_{h'}$ and the field $h'$ enters the counterterms only under spatial gradient. 

${\bf (4)}$ The counterterms $h'\partial_{t}h$ and ${v}'\partial_{t}{v}$ are allowed by the formal index $\delta$ but they are forbidden by the items {\bf (2)} and ${\bf (3)}$ because they do not contain a spatial gradient $\partial$. On the other hand, the Galilean
symmetry of the model requires, in particular, that the covariant derivatives $h'\nabla_{t}h$ and ${v}'\nabla_{t}{v}$ enter the counterterms as a single unit. Thus, counterterms $h'({v} \partial)h$ and ${v}'({v} \partial){v}$ are also forbidden.

Canonical dimensions analysis together with these considerations shows that superficial UV divergences are present in the following 1-irreducible Green’s functions: 
\[\langle {v_i}'{v_j}\rangle_{1-irr} \quad \mbox{with counterterm  }  \, {v_i}'{\partial^2}{v_j} \quad \left(\delta=2, \delta'=0\right)
\]
\[
\langle {v_i}'{v_j}'\rangle_{1-irr} \quad \mbox{with counterterm  } \, {v_i}'{\partial^2}{v_j}' \quad \left(\delta=2, \delta'=0\right)\\
\]
\[
\langle {h}'{h}^{n}\rangle_{1-irr} \quad \mbox{with counterterm  } \, {h}'{\partial^2}{h}^n \quad \left(\delta=2, \delta'=0\right)\\
\]
with any natural number $n$.

Inclusion of the corresponding counterterms can be reproduced by multiplicative renormalization of the fields and the parameters, i.e., the model (\ref{Action}), (\ref{NSAction}) is multiplicatively renormalizable. Furthermore, it appears that renormalization of the fields $h, h', v, v'$  is not required: their renormalized counterparts are given by the fields themselves.

Note that if functional $V(h)$ involves only odd powers of field $h$, the model acquires additional symmetry $h\rightarrow -h$, $h'\rightarrow -h'$ that forbids counterterms with even powers of $h$. Thus, such a model appears to be self-contained or ``closed with respect to the renormalization''.

\subsection{Renormalized action functionals} \label{sec:NS}

The renormalized action functional ${\cal S}_{{v} R}$ for the NS model in $d=2-\varepsilon$ 
has the following form:
\begin{equation}
{\cal S}_{{v} R} = \frac{1}{2} {{v}}' \left[g{\nu}^{3}\mu^{y}k^{2 + \varepsilon - y} + Z_{ii} g'{\nu}^{3}\mu^{\varepsilon}k^{2}\right]  {{v}}' + {v}' \left[ - \nabla_t  {{v}} + Z_{i}{\nu}\partial^2 {v}\right],
 \label{NSActionR}
\end{equation}
where we denoted
\begin{equation}
D_0 = D= g \nu^{3}\mu^{y}, \quad D'_0 = D' = g' \nu^{3}\mu^{\varepsilon}.
\label{NewCouplings}
\end{equation}
Here the renormalization mass $\mu$ is introduced as an additional parameter of the renormalized theory.  
These expressions follow from the fact that fields $h', h, {v}', {v}$ are
not renormalized.

The renormalization constants $Z_{i}$ and $Z_{ii}$ are chosen to eliminate the UV divergences of Green's functions $\langle{v}'{v}\rangle$ and $\langle{v}'{v}'\rangle$ respectively. 
In practical calculations, we will employ the minimal subtraction (MS) renormalization scheme, where these constants have the forms of pure poles in $y$, $\varepsilon$ and, in general, their linear combinations.

Renormalized action functional 
${\cal S}_{R} = {\cal S}_{h R} + {\cal S}_{{v} R}$  for the full model has the form
\begin{equation}
{\cal S}_{R}(\Phi) = \frac{1}{2}h'h' + h'\left\{- \nabla_{t}h + {\partial}^{2}V_{R}\left(h\right)\right\} 
+ {\cal S}_{{v} R}({\Phi}) ,
\label{ActionR}
\end{equation}
where $V_R(h)$ is the renormalized counterpart of the infinite series from definition (\ref{V}):
\begin{equation}
V_R(h) = \sum_{n=1}^{\infty}\, \frac{1}{n!}\,Z_{n} \lambda_{n}\, h^n.
\label{VR}
\end{equation}
The functional ${\cal S}_{{v} R}$ is given by expression (\ref{NSActionR}).
The renormalized action functional (\ref{ActionR}) is obtained from expression (\ref{Action}) using the following relations:
\begin{equation}
   \lambda_{n0} = \lambda_{n} Z_{n}, \quad
   \nu_{0} = \nu Z_{\nu}, \quad
   g'_{0} = g'{\mu}^{\varepsilon}Z_{g'}, \quad
   g_{n0} = g_{n}{\mu}^{\left(n - 1\right)\varepsilon/2}Z_{g_{n}},
   \label{multy}
\end{equation}
\[
 g_{0} = g{\mu}^{y}Z_{g}, \quad
 Z_{h} = Z_{h'} = Z_{{v}} = Z_{{v}'} = 1.
\]
The constants $Z_n$, $Z_i$ and $Z_{ii}$ are calculated directly from the diagrams for the Green's functions, and the rest are found from the relations
\begin{equation}
Z_{\nu}=Z_{i}, \quad
   Z_{g_{n}} = Z_n\, Z_{i}^{-(n+1)/2}, \quad
   Z_{g} = Z_{i}^{-3}, \quad
   Z_{g'} = Z_{ii}\, Z_{i}^{-3}. 
   \label{multy1}
\end{equation}
All of them follow from the definitions (\ref{NewCouplings}) and (\ref{multy}).

The use of renormalization schemes in the study of NS equation near $d=2$ is a complicated issue; see the work
\cite{AntiRonis} where the MS scheme and the Speer's analytical regularization scheme were compared. In this paper we employ the MS scheme which is unambiguous in the one-loop approximation.

\section{Calculation of the one-loop counterterm} \label{sec:one-loop}

The one-loop calculation of the renormalization constants $Z_i$ and $Z_{ii}$ in MS scheme yields the following expressions; cf.~equations (3.115) in~\cite{Red}: 
\begin{equation}
Z_{i} = 1 - \frac{g}{32\pi y} - \frac{g'}{32\pi\varepsilon} + \dots,
   \label{NSZ1cal}
\end{equation}
\begin{equation}
Z_{ii} = 1 - \frac{g^{2}}{32\pi g'\left(2y - \varepsilon\right)} - \frac{g}{16\pi y} - \frac{g'}{32\pi\varepsilon} + \dots\ .
   \label{NSZ2cal}
\end{equation}
Note that at $g=0$ one obtains $Z_{i}=Z_{ii}$. This result is exact to all orders in $g'$; see~\cite{FNS1,FNS}, sec.~3.10 in~\cite{Red} and references therein.

Now let us obtain the one-loop approximations for the constants $Z_{n}$ by explicitly constructing a closed form of the one-loop counterterm in the renormalized action (\ref{ActionR}). To this end, consider the expansion of the renormalized 1-irreducible Green’s functions generating functional 
($p$ is the number of loops):
\begin{equation}
    \Gamma_R (\Phi) = \sum_{p = 0}^{\infty} \, \Gamma^{(p)}(\Phi), \quad
    \Gamma^{(0)}(\Phi) = {\cal S}_{R}(\Phi).
    \label{Loops}
\end{equation}
The loopless (tree-like) term is just the renormalized action (\ref{ActionR}), while the one-loop contribution is given by
\begin{equation}
     \Gamma^{(1)}(\Phi) = - \frac{1}{2}\, {\rm Tr}\, {\left\{\ln{\frac{W}{W_0}}\right\}},
    \label{GammaOne}
\end{equation}
where $W$ is a linear operation defined by its kernel
\begin{equation}
    W(x,x') = - \frac{\delta^2 {\cal S}_{R}(\Phi)}{{\delta \Phi(x)}{\delta \Phi(x')}},
    \label{W}
\end{equation}
while $W_0$ is its counterpart for the free part of the action, i.e., the matrix $W_0^{-1}$ contains propagators for the theory (\ref{ActionR}). Both $W$ and $W_0$ are $4\times4$ matrices in the full set of fields $\Phi = \{h', h, {v}', {v}\}$.

Let us employ the MS scheme, put $Z_n=1$ in expression (\ref{GammaOne}) and take $Z_n$ in the first non-trivial order in the couplings $g$, $g'$ and $g_n$ in the loopless contribution (\ref{ActionR}). Then the one-loop approximations for the constants $Z_n$ can be found from expression (\ref{Loops}), where these constants cancel the UV divergences.

As the divergent part of expression (\ref{GammaOne}) does not contain the fields $v$ or $v'$, it is sufficient to consider the matrix $W$ at $v=v'=0$. Then, the matrix $W$ can be symbolically represented as  
\begin{equation} W=\left( \begin{array}{cccc}
-\partial^2h'\cdot V'' & L^T & - \partial h' & 0\\
L & -1 & \partial h & 0\\
h'\partial & - h\partial & 0 & M^T\\
0 & 0 & M & - D_f 
\label{Matrix}
\end{array} \right)\end{equation} 
Here, we omitted the vector indices for brevity; we also dropped the Dirac's function $\delta(x-x')$ that appears in the process of functional derivation and upon which all operators in the matrix $W$ act (with the exception of the element $W^{(h h)}$ where the dot stands to show how the derivative works). 
Notations $V'$ and $V''$ stand for the derivatives of the series $V$ (obtained from $V_R (h)$ by assuming $Z_n = 1$) with respect to the field $h = h(x)$ as if $V$ were a function of a single variable $h$.
The correlation function $D_f$ is taken from (\ref{noiseNS}) and (\ref{noiseNS1}) while $L = \partial_t - \partial^2 V'$, $M = \partial_t - \nu\partial^2$ and transposed counterparts are $L^T = - \partial_t - V'\partial^2$, $M^T = - \partial_t - \nu\partial^2$.

To obtain explicit form of the constants $Z_n$, only the divergent part of the expression (\ref{GammaOne}) is required, which has the form 
\begin{equation}
    \int{dx \partial^2 h'(x) R(h(x))},
    \label{DivPart}
\end{equation}
that involves function $R(h)$ similar to $V(h)$. Hence it follows that it is sufficient to know the expression (\ref{GammaOne}) up to the first order with respect to the elements $W^{(h h)}$, $W^{(h {v})}$ and $W^{({v} h)}$ of the matrix (\ref{Matrix}) that are linear in $h'$. To extricate that part, one can use functional variations. Let us decompose the matrix $W$ as follows:
\begin{equation}W=W_0 + \delta W, 
\end{equation}
then the logarithm $\ln [W/W_0]$ in the expression (\ref{GammaOne}) can be expanded as
\begin{eqnarray}
{\rm Tr}\,
{\left\{\ln \frac{W}{W_0}\right\}}&=& {\rm Tr}\,{\left\{\ln \left[1+W_0^{-1}\delta W\right]\right\}} \nonumber\\
&=&{\rm Tr}\,{\left\{W_0^{-1}\delta W - \frac{1}{2}\, W_0^{-1}\delta W\, W_0^{-1}\delta W + \dots \right\}}.
\end{eqnarray}

The divergent part of the needed form is contained in the first two terms of this expansion
and can be expressed as ${\rm Tr}\,{\ln [W/W_0]} \simeq - I_1 + 2I_2$, where

\begin{equation}
      I_{1} = \int{dx\, D^{(h h)}\left(x, x\right)V''(h(x)){\partial}^{2}h'(x)}
      \label{I1}
\end{equation}
and
\begin{equation}
      I_{2} = {\int} dx{\int}{dx' \, {\partial}_{i}h(x)
      D^{(h' h)}\left(x, x'\right)D^{({{v}} {{v}})}_{ij}\left(x, x'\right){\partial}_{j}h'(x')}.       
      \label{I2}
\end{equation}
Here $D^{(\Phi \Phi)}$ are the corresponding elements of the matrix $W^{-1}$ for zero values of the fields ${v}$, ${v}'$ and $h'$. By its very meaning, $D^{(h h)}$ is the propagator $\langle h h\rangle_0$ for the model (\ref{ActionR}) with $Z_n = 1$ and with $\nu\partial^2$ substituted by $\partial^2 V'$ in the denominator. Similarly, $D^{(h' h)}$ is the propagator $\langle h' h\rangle_0$ with the same substitution (the so-called response function in the external field), while $D_{ij}^{({v} {v})}$ is just the correlation function (\ref{noiseNS}).

The external fields $h(x)$ and $h'(x)$ in the expressions (\ref{I1}) and (\ref{I2}) stand under derivatives, which means that the rest of the integrands diverge only nearly logarithmically. Thus, all the external frequencies and momenta can be set to zero during the calculation of the divergent parts of these integrals. This is possible because the IR regularization is provided by the cut off parameter $m$ from the expression (\ref{noiseNS}). Additionally, the factors ${\partial}^{2}h'(x)$, ${\partial}_{i}h(x)$, ${\partial}_{j}h'(x')$ as well as the field $h$ in the expressions (\ref{I1}) and (\ref{I2}) can be treated as constants.

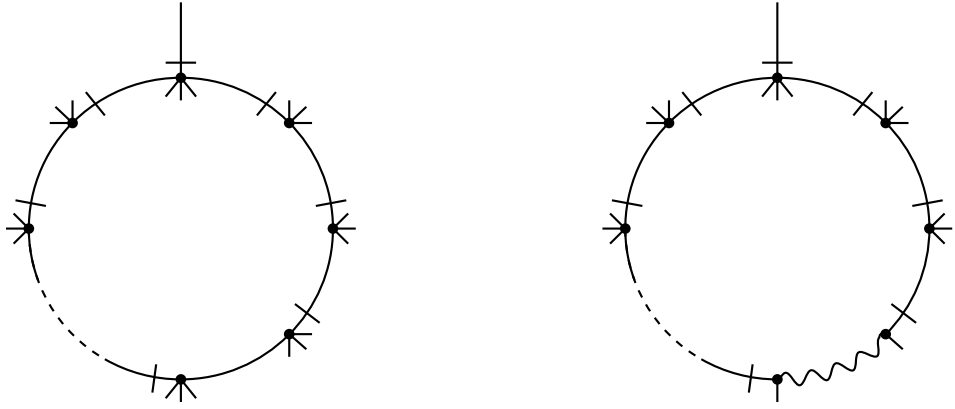
\begin{figure}[h!]
\usetikzlibrary{decorations.pathmorphing}
\tikzset{snake it/.style={decorate, decoration=snake}}
\minipage{0.4\textwidth}
\begin{center}
\begin{tikzpicture}
   \draw[thick] (0,0) arc (0:200:2);
   \draw[dashed, thick] (-4,0) arc (180:240:2);
   \draw[thick] (0,0) arc (0:-120:2);
   \draw[thick] (-2,2) - - (-2,3);
   \fill[black] (-2,2) circle(2pt);
   \fill[black] (0,0) circle(2pt);  
   \fill[black] (-4,0) circle(2pt);  
   \fill[black] (-2,-2) circle(2pt);
   \fill[black] (-0.575,1.4) circle(2pt);
   \fill[black] (-0.575,-1.4) circle(2pt);
 \fill[black] (-3.425,1.4) circle(2pt);
   \draw[thick] (-2,2) - - (-2,1.7);
   \draw[thick] (-2,2) - - (-2.2,1.75);
   \draw[thick] (-2,2) - - (-1.8,1.75);
   \draw[thick] (-2.2,2.2) - - (-1.8,2.2);
   \draw[thick] (0,0) - - (0.3,0);
   \draw[thick] (0,0) - - (0.2,0.2);
   \draw[thick] (0,0) - - (0.2,-0.2);
   \draw[thick] (-4,0) - - (-4.3,0);
   \draw[thick] (-4,0) - - (-4.2,0.2);
   \draw[thick] (-4,0) - - (-4.2,-0.2);
    \draw[thick] (-2,-2) - - (-2,-2.3);
   \draw[thick] (-2,-2) - - (-2.2,-2.25);
   \draw[thick] (-2,-2) - - (-1.8,-2.25);
   \draw[thick](-0.575,1.4) - - (-0.35,1.615);
   \draw[thick](-0.575,1.4) - - (-0.575,1.7);
   \draw[thick] (-0.575,1.4)- - (-0.275,1.4);
    \draw[thick] (-3.425,1.4)  - - (-3.65,1.615);
   \draw[thick] (-3.425,1.4)  - - (-3.425,1.7);
   \draw[thick]  (-3.425,1.4) - - (-3.725,1.4);
   \draw[thick](-0.575,-1.4) - - (-0.35,-1.6);
   \draw[thick](-0.575,-1.4) - - (-0.575,-1.7);
   \draw[thick] (-0.575,-1.4)- - (-0.275,-1.4);
   \draw[thick] (-1,1.5) - - (-0.75,1.81);
   \draw[thick] (-0.225,0.3) - - (0.175,0.375);
   \draw[thick] (-0.5,-1) - - (-0.175,-1.25);
   \draw[thick](-3,1.5) - - (-3.25,1.81);
   \draw[thick](-3.775,0.3) - -(-4.175,0.375);
   \draw[thick] (-2.375,-2.175) - - (-2.325,-1.8);
\end{tikzpicture}
\end{center}
\endminipage
\hfill
\minipage{0.4\textwidth}
\usetikzlibrary{decorations.pathmorphing}
\begin{center}
\begin{tikzpicture}
\usetikzlibrary{decorations.pathmorphing}
\tikzset{snake it/.style={decorate, decoration=snake}}
 \draw[thick] (0,0) arc (0:200:2);
   \draw[dashed, thick] (-4,0) arc (180:240:2);
   \draw[thick, snake it] (-2,-2) arc (270:317:2);
   \draw[thick] (-2,-2) arc (-90:-120:2);
    \draw[thick] (0,0) arc (0:-45:2);
   \draw[thick] (-2,2) - - (-2,3);
   \fill[black] (-2,2) circle(2pt);
   \fill[black] (0,0) circle(2pt);  
   \fill[black] (-4,0) circle(2pt);  
   \fill[black] (-2,-2) circle(2pt);
   \fill[black] (-0.575,1.4) circle(2pt);
   \fill[black] (-0.575,-1.4) circle(2pt);
 \fill[black] (-3.425,1.4) circle(2pt);
   \draw[thick] (-2,2) - - (-2,1.7);
   \draw[thick] (-2,2) - - (-2.2,1.75);
   \draw[thick] (-2,2) - - (-1.8,1.75);
   \draw[thick] (-2.2,2.2) - - (-1.8,2.2);
   \draw[thick] (0,0) - - (0.3,0);
   \draw[thick] (0,0) - - (0.2,0.2);
   \draw[thick] (0,0) - - (0.2,-0.2);
   \draw[thick] (-4,0) - - (-4.3,0);
   \draw[thick] (-4,0) - - (-4.2,0.2);
   \draw[thick] (-4,0) - - (-4.2,-0.2);
    \draw[thick] (-2,-2) - - (-2,-2.3);
   \draw[thick](-0.575,1.4) - - (-0.35,1.615);
   \draw[thick](-0.575,1.4) - - (-0.575,1.7);
   \draw[thick] (-0.575,1.4)- - (-0.275,1.4);
    \draw[thick] (-3.425,1.4)  - - (-3.65,1.615);
   \draw[thick] (-3.425,1.4)  - - (-3.425,1.7);
   \draw[thick]  (-3.425,1.4) - - (-3.725,1.4);
   \draw[thick](-0.575,-1.4) - - (-0.35,-1.6);
   \draw[thick] (-1,1.5) - - (-0.75,1.81);
   \draw[thick] (-0.225,0.3) - - (0.175,0.375);
   \draw[thick] (-0.5,-1) - - (-0.175,-1.25);
   \draw[thick](-3,1.5) - - (-3.25,1.81);
   \draw[thick](-3.775,0.3) - -(-4.175,0.375);
   \draw[thick] (-2.375,-2.175) - - (-2.325,-1.8);
\end{tikzpicture}
\end{center}
\endminipage
\caption{Diagrammatic representations for integrals $I_1$ (on the left) and $I_2$ (on the right), see diagrammatic technique (\ref{D1})~-- (\ref{D2}) and Figure \ref{FeynmanVert}. No frequencies or momenta flow from the external tails of the diagrams, the only integration is over momentum and frequency circulating inside the loops. Thus, the heterogeneity of the fields $h(x)$ and $h'(x)$ can be neglected.}
\label{PavlikDiagrams}
\end{figure}
So the integrals are easily calculated by going over to the Fourier
(frequency-momentum) representation.

First, let us carry out intermediate calculations of $D^{(h h)}$:
 \begin{eqnarray}
D^{(h h)}(x,x) &=&
\int_{k>m} \frac{d{\bf k}}{(2\pi)^d}\, 
\int \frac{d\omega}{2\pi}\,
\frac{1}{\omega^2+[k^2\,V'(h)]^2}=
 \nonumber \\  &=&
\frac{1}{2V'(h)}
\int_{k>m} \frac{d{\bf k}}{(2\pi)^d}\,
\frac{1}{k^2} = 
 \nonumber \\  &=&
\frac{S_d}{(2\pi)^d}\,\frac{1}{2V'(h)}\,
\int_{m}^{\infty}\, dk\, k^{d-3} =
\frac{S_d}{(2\pi)^d}\,\frac{1}{2V'(h)}\, \frac{1}{\varepsilon}\, m^{-\varepsilon},
\label{Forintegral1}
\end{eqnarray} 
where it was taken into account that $d = 2-\varepsilon$ and where $S_d =2\pi^{d/2}/\Gamma(d/2)$ is the area of the unit sphere in the $d$-dimensional space. Then for the integral (\ref{I1}), one obtains:
\begin{equation}
      I_{1} = \frac{S_d}{2(2\pi)^d} \, \left(\frac{\mu}{m}\right)^{\varepsilon} \frac{1}{\varepsilon}\int{dx\, F_{1}\left(h(x)\right)\, {\partial}^{2}h'(x)},
      \label{Integral1}
\end{equation}
where function $F_{1}\left(h(x)\right)$ is defined as 
\begin{equation}
F_{1}\left(h\right) = \mu^{-\varepsilon}\, \frac{V''(h)}{V'(h)}.
\label{F1}
\end{equation}
For the integral entering (\ref{I2}), one has:
\begin{eqnarray}
& \int {dx'} &  D^{(h' h)}(x,x')\,
D_{ij}^{({v}{v})}(x,x') =
 \nonumber \\  =
& \int_{k>m} & \frac{d{\bf k}}{(2\pi)^d}\, 
\int \frac{d\omega}{2\pi}\,
\frac{P_{ij}\, D_f(k) }{{\rm i}\omega+k^2\, V'(h)} \cdot 
\frac{1}{\omega^2 + \nu^2\, k^4} , \hspace{1cm}
\label{Forintegral2}
\end{eqnarray} 
where the function $D_f(k)$ is defined by the expression (\ref{noiseNS1}).
The integral (\ref{Forintegral2}) over the frequency $\omega$ can be calculated straightforwardly:
\begin{eqnarray}
& {\int dx'} &\, D^{(h' h)}(x,x')\,
D_{ij}^{({v}{v})}(x,x') =  \frac{1}{2}
\int_{k>m} \frac{d{\bf k}}{(2\pi)^d}\, \frac{P_{ij}\, D_f(k)}{\nu k^4 (\nu+V')} =
\nonumber \\ &=& 
\!\!\!\!\!\!\!
 \frac{d-1}{2d}\, \delta_{ij}\, \frac{S_d}{(2\pi)^d}\, \frac{\nu^2}{\nu+V'}\left\{\frac{g}{y}\, \left(\frac{\mu}{m}\right)^y + \frac{g'}{\varepsilon}\, \left(\frac{\mu}{m}\right)^\varepsilon\right\}. \hspace{1cm}
\label{ForI2noOmega}
\end{eqnarray} 
Thus, we obtain the following expression for the integral (\ref{I2}):
\begin{equation}
      I_{2} = - \frac{(d-1)}{2d}\, \frac{S_d}{(2\pi)^d}\,\left\{\frac{g}{y}\, \left(\frac{\mu}{m}\right)^y + \frac{g'}{\varepsilon}\, \left(\frac{\mu}{m}\right)^\varepsilon\right\} \int{dx\, F_2\left(h(x)\right)\, \partial^2 h'(x)}, \hspace{1cm}
      \label{Integral2}
\end{equation}
where the function $F_{2}(h)$ can be expressed as follows:
\begin{equation}
    F_{2}\left(h\right) = \int_0^{h}{d{\tilde h}\, \frac{\nu^2}{\nu+V'\left({\tilde h}\right)}}.
    \label{F2}
\end{equation}
The lower limit of integration (\ref{F2}) can in fact be chosen arbitrarily; see the explanation below. Here, we used the integration by parts and the chain rule $\partial F_2/\partial x= (\partial F_2/\partial h) (\partial h/\partial x)$ with $\partial F_2/\partial h={\nu^2}/{(\nu+V')}$.

Then, taking into account the results (\ref{Integral1}) and (\ref{Integral2}), we obtain for the divergent part of the one-loop contribution the following expression: 
\begin{equation}
     \Gamma^{(1)}(\Phi) = \frac{a_{1}}{\varepsilon}\,
     \left(\frac{\mu}{m}\right)^{\varepsilon}\,
      {\int}dx\,
     h'(x)\, {\partial}^{2}\,
     F_{1}\left(h(x)\right) + 
     \label{Ratio}     
\end{equation}
     \[
     +\,  a_2
     \left\{\frac{g}{y}\, \left(\frac{\mu}{m}\right)^y + \frac{g'}{\varepsilon}\, \left(\frac{\mu}{m}\right)^\varepsilon\right\}\,
      {\int}dx\,
     h'(x)\, {\partial}^{2}\,
     F_{2}\left(h(x)\right),
     \]
where we denoted: 
\begin{equation}
    a_1 = \frac{S_d}{4(2\pi)^d}. \hspace{1cm}
    a_2 = \frac{d-1}{2d}\,\frac{S_d}{(2\pi)^d}; 
    \label{A}
\end{equation}
note that $a_1 = a_2 = 1/(8\pi)$ for $d=2$.

It would be preferable to continue the renormalization analysis in terms of closed functional representations like (\ref{F1}), (\ref{F2})
and (\ref{Ratio}). Unfortunately, presently the authors know of no way to perform it. Hopefully,
the functional RG might be better suited for this end (see more on this in Sec.~8).

Thus, for lack of a better solution, here we expand the functions $F_1\left(h\right)$ and $F_2\left(h\right)$ in powers of the field $h$ again and arrive at the series of the form (\ref{V}) and (\ref{VR}):
\begin{equation}
    F_{1}\left(h\right) = \sum_{n=0}^{\infty}\frac{1}{n!}\,
    {\mu}^{\varepsilon(n-1)/2}\,
    {\nu}^{\left(n+1\right)/2} \, r_{n}\,h^{n}
    \label{F1Series}    
\end{equation}
and
\begin{equation}
    F_{2}\left(h\right) = \sum_{n=0}^{\infty}\frac{1}{n!}\,
    {\mu}^{\varepsilon(n-1)/2}\,
    {\nu}^{\left(n+1\right)/2} \, s_{n}\,h^{n},
\label{F2Series}    
\end{equation}
with dimensionless coefficients $r_n$ and $s_n$. By using the expression (\ref{VR}) with the substitution $Z_{n}=1$, one can express them in terms of the couplings $g_n$ for each order in $n$. We list the first four of these coefficients below to illustrate their general structure (recall that in the loop expansion one assumes $g_n \simeq g_2^{(n-1)}$):
\begin{eqnarray}
r_1 &=& \frac{g_3}{g_1} - \left(\frac{g_2}{g_1}\right)^2,  \nonumber \\
r_2 &=& \frac{g_4}{g_1} - 3 \,\frac{g_3\, g_2}{g^2_1} + 
2 \left(\frac{g_2}{g_1}\right)^3 , 
 \nonumber \\  
r_3 &=& \frac{g_5}{g_1} - 4 \,\frac{g_4 \, g_2}{g^2_1}- 3\,\left(\frac{g_3}{g_1}\right)^2 + 12\,\frac{g_3\, g_2^2}{g^3_1} -6\,\left(\frac{g_2}{g_1}\right)^4,
 \nonumber \\
r_4 &=& \frac{g_6}{g_1} - 5\,\frac{g_5\, g_2}{g^2_1} + 20 \,\frac{g_4 \,g_2^2}{g^3_1} -10 \,\frac{g_4\, g_3}{g^2_1} +  \nonumber \\
&+& 30\, \frac{g_3^2 \,g_2}{g^3_1}  -60 \,\frac{g_3 \,g_2^{3}}{g^4_1} + 24 \,\left(\frac{g_2}{g_1}\right)^5
 \label{Rn}  
\end{eqnarray}
and 
\begin{eqnarray}
s_1 &=& \frac{1}{(g_1 + 1)}, \quad
s_2 = \frac{-g_2}{\left(g_1 + 1\right)^2}, \quad
s_3 = \frac{-g_3}{\left(g_1 + 1\right)^2} + \frac{2g^2_2}{\left(g_1+1\right)^3}.
\label{Qn}  
\end{eqnarray}
The coefficients $r_0$ and $s_0$ are not shown because they do not enter the expression (\ref{Ratio}) for the divergent part of the one-loop contribution. For the same reason, the lower limit of integration in expression (\ref{F2}) responsible for the coefficient $s_0$ can be chosen arbitrarily.

Now using the fact that the poles in $y$ and $\varepsilon$ are cancelled in the sum of the first two terms in expression (\ref{Loops}), the explicit expressions for the renormalization constants can be derived. In the MS scheme, these constants have the forms ``$Z_n = 1+$ only pure poles in $y$ and $\varepsilon$,'' so that the factors such as $\left(\mu/m\right)^y$ and $\left(\mu/m\right)^{\varepsilon}$ should be replaced by unity. The final expression for the renormalization constants are:
\begin{equation}
    Z_{n} = 1 - \frac{1}{8\pi\varepsilon}\,\frac{r_n}{ g_n} - g\,\frac{1}{8\pi y}\,\frac{s_n}{g_n} - g'\,\frac{1}{8\pi\varepsilon}\,\frac{s_n}{g_n} + \dots
    \label{Zn}
\end{equation}
with the coefficients $r_n$, $s_n$ from (\ref{F1Series}), (\ref{F2Series}) and the factors $1/8\pi$ from~(\ref{A}) for~$d=2$.

\section{RG equations and RG functions} \label{sec:RGE}

For a multiplicatively renormalizable model the RG equations
can be derived in a standard fashion; see, e.g. \cite{Vasiliev}. For the 
model (\ref{Action}), (\ref{NSAction}) one obtains:
\begin{equation}
\left\{ {\cal D}_{\mu} - \beta_{g} \partial_g - \beta_{g'} \partial_{g'} - \sum_{n=1}^{\infty}\, \beta_n \partial_{g_n} - \gamma_{\nu}  {\cal D}_{\nu}
+ \gamma_{G} \right\}\, G(e;\dots)=0.
\label{RGE}
\end{equation}
Here $G(\cdot)$ is a certain renormalized (and hence UV finite) Green's function of the model (\ref{ActionR}) expressed in terms of the full set of renormalized variables $e=\left\{g,g',g_n,\nu,\mu,m \right\}$. The ellipsis denotes other variables such as coordinates (or momenta) and times (or frequencies).
Here and below $\partial_x = \partial/\partial x$ and ${\cal D}_x = x \partial_x$ for any variable $x$.

The coefficients in the RG differential operator (\ref{RGE}) are
the anomalous dimensions $\gamma$ and the $\beta$ functions defined as
\begin{equation}
\gamma_{F}= \widetilde{\cal D}_{\mu}\, \ln Z_{F} \quad
{\rm for\, any\,} F,
\quad
\beta_g = \widetilde{\cal D}_{\mu}\,g, \quad
\beta_{g'} = \widetilde{\cal D}_{\mu}\,{g'}, \quad
\beta_n = \widetilde{\cal D}_{\mu}\, g_n , \hspace{1cm}
\label{RGF}
\end{equation}
where $\widetilde{\cal D}_{\mu}$ is the differential operation
${\cal D}_{\mu}$ at fixed non-renormalized (bare) parameters.

The definitions (\ref{RGF}) and relations (\ref{multy}), (\ref{multy1}) allow one to express all those RG functions in terms of the anomalous dimensions $\gamma_{i}$, $\gamma_{ii}$ and $\gamma_n$ as follows:
\begin{eqnarray}
\gamma_{g} &=& -3 \gamma_i,  \quad
\gamma_{g'} = \gamma_{ii} -3 \gamma_i, \quad
\gamma_{g_n} =  \gamma_n - (n+1){\gamma_i}/2, 
 \nonumber \\
\gamma_{\nu} &=& \gamma_i, \quad \gamma_{\Phi}=0 \ {\rm for\ all}\ 
\Phi
\label{relations} 
\end{eqnarray}
and
\begin{eqnarray}
\beta_g &=& g\left[-y -\gamma_g\right] = g\left[-y + 3\gamma_i\right], \nonumber \\
\beta_{g'} &=& g'\left[-\varepsilon -\gamma_{g'}\right] = g'\left[-\varepsilon + 3\gamma_i - {\gamma}_{ii}\right],
\nonumber \\
\beta_n &=& g_n \left[-(n-1){\varepsilon}/2 - \gamma_{g_n}\right]=
g_n\left[-(n-1){\varepsilon}/2-\gamma_n +(n+1){\gamma_i}/2\right]. 
\label{relations1} 
\end{eqnarray}
Here the operation $\widetilde{\cal D}_{\mu}$ acts on the functions that depend only on couplings and takes on the form:
\begin{equation}
\widetilde{\cal D}_{\mu} = \beta_g \partial_g + \beta_{g'} \partial_{g'} + \sum_{n=1}^{\infty} \, \beta_n \partial_{g_n},
\label{derevo}
\end{equation}
which can be reduced with the required accuracy to
\begin{equation}
\widetilde{\cal D}_{\mu} \simeq -y {\cal D}_{g} - \varepsilon {\cal D}_{g'} -
(\varepsilon/2){\cal D}, \quad {\rm where}\ \ {\cal D}=\sum_{n=1}^{\infty}\, (n-1)\, {\cal D}_{g_n}.
\label{derevo1}
\end{equation}
The following relations are directly checked:
\begin{eqnarray}
    {\cal D}\,{r_n} = (n+1){r_n},  \quad  
    {\cal D}\, \left({r_n}/{g_n}\right) = 2 \, \left({r_n}/{g_n}\right),  \quad
    {\cal D}\, \left({s_n}/{g_n}\right) = {0}. 
    \label{equalities}
\end{eqnarray}

They are naturally interpreted as homogeneity relations with the assumption that \mbox{$g_n \sim g_2^{(n-1)}$}: then $r_n \sim g_2^{(n+1)}$ and $s_n \sim g_2^n$.
Also note that $g_1\sim 1$ and ${\cal D}_{g_1}$ does not contribute to ${\cal D}$ in (\ref{derevo1}).

These relations
along with the explicit expressions of the constants $Z_{i}$, $Z_{ii}$ and $Z_{n}$ in~(\ref{NSZ1cal}), (\ref{NSZ2cal}) and (\ref{Zn}) 
give the following one-loops result for the corresponding
anomalous dimensions:
\begin{eqnarray}
    \gamma_i &=& \frac{1}{32\pi}\left(g + g'\right)+\dots, 
    \label{gammas} \\
    \gamma_{ii} &=&  \frac{1}{32{\pi}g'}\left(g + g'\right)^2+\dots, 
     \label{gammas1} \\
     \gamma_n &=& \frac{1}{8\pi g_n}\, \left\{ r_n +  s_n (g+g')\right\}+\dots,
    \label{gamman} 
\end{eqnarray}   
with the coefficients $r_n$ and $s_n$ from (\ref{Rn}) and (\ref{Qn}), respectively.

Using the relations (\ref{relations}) and (\ref{gammas}), one obtains for the functions $\beta_{g}$, $\beta_{g'}$ and $\beta_{g_n}$ in one-loop approximation:
\begin{eqnarray}
    \beta_g &=& g\left[-y + \frac{3}{32\pi}\left(g + g'\right) +\dots \right], \label{betag} \\
\beta_{g'} &=& g'\left[-\varepsilon + \frac{3}{32\pi}\left(g + g'\right) - \frac{1}{32{\pi}g'}\left(g + g'\right)^2 
 + \dots\right], 
 \label{beta2}\\
\beta_{g_n}&=&g_n\left[-\frac{(n-1)\varepsilon}{2}-\gamma_n+\frac{(n+1)}{2}\frac{1}{32\pi}\left(g + g'\right)+ \dots\right]. 
\label{betan}
\end{eqnarray}
Of course, the expressions (\ref{gammas}), (\ref{gammas1}) and (\ref{betag}), (\ref{beta2}) are in agreement (up to notation) with those presented in 
section 3.10~in~\cite{Red} and in \cite{HoNa,AntiRonis} for the pure NS model.

\section{Attractors of the RG equations, scaling behaviour and exact critical dimensions} 
\label{sec:scaling}

It is well known that the IR asymptotic behaviour of the Green’s functions of a multiplicatively renormalizable model is governed by IR attractive fixed points of the corresponding RG equations. Those points are specified in the parameter space of renormalized couplings and are determined from the requirement that all the $\beta$ functions vanish.
Since our model (\ref{Action}), (\ref{NSAction}) includes an infinite set of couplings $g,g',g_{n}$, the corresponding space is infinite-dimensional.
The fixed points are found from the requirement that 
\begin{equation}
    \beta_g \left(g^*, {g'}^*\right) = 0, \quad
    \beta_{g'} \left(g^*, {g'}^*\right) = 0, \quad
    \beta_{g_n} \left(g^*, {g'}^*,{g_n}^*\right) = 0 \quad   (n\ge1).
    \label{Attractbetas}
\end{equation} 

The equations (\ref{Attractbetas}) for the functions (\ref{betag}), (\ref{beta2}) 
have three possible solutions, which can be written with one-loop accuracy in the form:
\begin{eqnarray}
    (1) \quad g^* &=& 0 \, , \quad 
    g'^* = 0,
    \label{IRGS}\\ 
    (2) \quad g^* &=& 0\, , \quad 
    g'^* = 16{\pi}{\varepsilon}+\dots, 
    \label{IRGS2}\\
    (3) \quad g^* &=& \frac{32\pi}{9} \, \frac{y\, (2y - 3\varepsilon)}{(y - \varepsilon)} \, +\dots, \quad g'^* = \frac{32\pi}{9} \frac{y^2}{(y - \varepsilon)} +\dots.
    \label{IRGS3}
\end{eqnarray}
The first point corresponds to the free (non-interacting) theory and is IR attractive for $\varepsilon<0$, $y<0$. The second point, attractive for $\varepsilon>0$, $y-3\varepsilon/2<0$, corresponds to a fluid in thermal equilibrium. The third point, attractive for $y>0$, $y-3\varepsilon/2>0$, corresponds to a strongly turbulent fluid. Note that in  the regions of IR stability one has $g^*, g'^* \ge 0$, as required by physical considerations.

Of course, these results coincide, up to the notation, with the ones obtained in \cite{HoNa} for the pure NS equation\footnote{See the paragraph between equations (19) and (20) in \cite{HoNa} but note that the authors omitted strokes after $a$ and $g$ in the last expressions for $a_*$ and $g_*$. The eigenvalues that characterize stability of the points can also be found in \cite{HoNa}; see expression (21) and the paragraph above it.}; see also section~3.10 in the monograph~\cite{Red}.

Substitution of the solutions (\ref{IRGS})~-- (\ref{IRGS3})
into the remaining $\beta$ functions gives rise to three infinite sets of equations $\beta_n=0$ for the remaining parameters $g_n$ with $n\ge1$. The direct analysis shows that, for each case, the parameters $g_1^*$ and $g_2^*$ can be chosen arbitrarily, while $g_3^*$ is determined in a unique way from the equation $\beta_1=0$.
Then  $g_4^*$ is determined by the equation $\beta_2=0$ and so on:
sequential substitution of the solutions of 
$\beta_k = 0$ with $k \le n$ into the remaining equations $\beta_j = 0$ with $j > n$ results in expressions for all the parameters $g^*_n$ in terms of two coordinates $g^*_1$ and $g^*_2$. 
This picture is a consequence of the fact that the coefficient $r_n$ entering the function $\beta_n$ 
involves coupling constants $g_{k}$ with $k$ running up to $(n+2)$ but only $g_{n+2}$ is not encountered in coefficients of the previous order; see expression (\ref{Rn}).
The coefficients $s_n$ from (\ref{Qn}) do not involve $g_j$ with $j>n+1$ and therefore do not revise these considerations.

Thus, we conclude that attractors of the RG equation (\ref{RGE}) have the form of three two-dimensional surfaces in the infinite-dimensional space of coupling constants~$g^*$, $g'^*$ and $g^*_n$.

In general, the character of a fixed point is determined by the eigenvalues of the 
stability matrix $\partial\beta/\partial g$ that involves the full set of couplings and 
$\beta$ functions: 
the point is IR attractive if the real parts of all its eigenvalues are positive; see, e.g. Sec.~1.42 in  \cite{Vasiliev}.
 However, in the present model, the stability matrix is semi-infinite so the required analysis appears rather embarrassing. We only managed to check that the trace of the matrix is positive for certain areas on the attractors; cf.~\cite{ABK}. Although it is only a necessary condition of IR attractiveness (and rather weak due to the infinite number of terms),
it still leaves the possibility of IR regions existing on the attractor surfaces open.

If such regions indeed exist, the Green's functions of the model will exhibit scaling behaviour of the type (\ref{scaling}) with certain critical dimensions.
The critical dimensions of any quantity (a field or parameter) in dynamical models are determined by (see, e.g. Sections 5.16 and 6.7 in \cite{Vasiliev})  
\begin{equation}
    \Delta_F = d^k_F + \Delta_{\omega} d^\omega_F + \gamma^*_F \quad
    \Delta_\omega = 2 - \gamma^*_{\nu}.
    \label{Delta}
\end{equation}

For the first surface, the models (\ref{Action}) and (\ref{NSAction}) decouple and can be studied separately. The model (\ref{Action}) was studied in~\cite{AV}. It was shown that the critical dimensions are non-universal in the sense that they depend on the specific choice of a point on the only surface of fixed points. However, they obey the exact relations $\Delta_{h'}=d-\Delta_h$, $2\Delta_h =d-\Delta_{\omega}$~\cite{AV}. 

In its turn, the model (\ref{NSAction}) becomes effectively Gaussian and one easily obtains exact expressions $\Delta_v=1$, $\Delta_{v'}=(d-1)$ and $\Delta_{\omega}=2$. Thus, the dimensions $\Delta_{\omega}$ for the fields $h$ and $v$ appear in general different, the phenomenon sometimes referred to as ``weak scaling,'' see, e.g.~\cite{Folk}.

The situation becomes much more interesting for the two surfaces originated from the fixed points (\ref{IRGS2}) and (\ref{IRGS3}). From the definitions (\ref{RGF}), relations (\ref{relations1}) 
and equations $\beta_g=\beta_{g'}=0$, one obtains the exact expressions for the values $\gamma_{\nu}^*$ on the attractors: $\gamma_{\nu}^*=\varepsilon/2$ for the second surface and 
$\gamma_{\nu}^*=y/3$ for the third one.

Using the canonical dimensions from Table~1 and taking into account that $\gamma_\Phi = 0$,
one obtains resulting expressions of critical dimensions for each surface, presented in Table~2. 
\begin{table}[h!]
\caption{Critical dimensions for the model (\ref{Action}), (\ref{NSAction}).}
\label{t2}
     \begin{tabular}{c||c|c|c|c|c}
     \hline\hline
     Attractors & $\Delta_{\omega}$ & $\Delta_{h'}$ & $\Delta_{h}$ & $\Delta_{v'}$ & $\Delta_{v}$ \\ \hline
    $2^{\tt nd}$ Surface & $2 - \varepsilon/2$ & $2 - 3\varepsilon/4$ & $-\varepsilon/4$ & $1 - \varepsilon/2$ & $1 - \varepsilon/2$ \\ \hline
    $3^{\tt rd}$ Surface & $2 - y/3$ & $2 - (y+3\varepsilon)/6$ & $(y-3\varepsilon)/6$ & $1-\varepsilon+y/3$ & $1 - y/3$ \\ \hline\hline
\end{tabular}
\end{table}
All these  expressions are universal in the sense that they do not depend on the choice of the point on the attractor surface: they depend only on the spatial dimension $d$ for the second surface and on $d$ and the exponent $y$ for the third one. What is more,
although the RG functions are derived only in the one-loop approximation, these expressions are exact: they have no higher-order corrections in $\varepsilon$ and $y$. This is probably the most interesting result of the present study.

It remains to note that in the traditional notation $z=\Delta_{\omega}$
and $\chi=-\Delta_{h}$ in the scaling expressions like (\ref{scaling}).

\section{Conclusion and discussion} \label{sec:conc}

We studied the extended Pavlik’s model (\ref{noise}), (\ref{advection}), (\ref{V}) of kinetic roughening using the field theoretic RG and taking into account random motion of the environment. The latter was modelled by the stochastic differential Navier--Stokes equation (\ref{NSeq}) with the correlation function of the random force (\ref{noiseNS}) containing two terms (\ref{noiseNS1}) that allow to consider both a fluid in thermal equilibrium and a turbulent fluid.

We established multiplicative renormalizability of the corresponding field theoretic model~(\ref{Action}), (\ref{NSAction}) in the infinite space of coupling constants and constructed the renormalized action functional~(\ref{ActionR}). 
Then we found the one-loop counterterm in an explicit closed form (\ref{Ratio}) and the full set of the RG functions (\ref{relations1}), (\ref{gammas})~-- (\ref{gamman}) in the one-loop approximation.

Analysis of attractors of the RG equation (\ref{RGE}) revealed three two-dimensional surfaces of fixed points~(\ref{IRGS})~-- (\ref{IRGS3}).

The first surface (\ref{IRGS}) corresponds to the situation when the dynamics of the height
field $h$ and the velocity field $v$ decouple (in the leading order of the IR asymptotic behaviour). Thus, according to the results derived  in \cite{AV} for the pure extended Pavlik's model, the critical dimensions are non-universal in the sense that they depend on the choice of the point on the surface. However, they satisfy the exact expressions 
$\Delta_{h'}=d-\Delta_h$, $2\Delta_h =d-\Delta_{\omega}$; see~\cite{AV}.

The surfaces~(\ref{IRGS2}) and (\ref{IRGS3}) correspond to regimes where both the kinetic roughening and the random environment motion are relevant at the same time. The first case corresponds to a fluid in thermal equilibrium, while the second one deals with a turbulent fluid. For the both cases, the critical dimensions are found exactly and they are universal; see~Table~2.

In quantum field theory, the models that require infinitely many counterterms  
traditionally faced sceptical attitudes for being non-renormalizable and therefore for having no predictive power. However, now the common opinion is changing, although a generally accepted interpretation is still not achieved, in particular, because of a wide variety and diversification of such models in comparison with usual renormalizable ones.\footnote{All happy families are alike; each unhappy family is unhappy in its own way (``Anna Karenina'' by Leo Tolstoy).}
For a recent discussion see, e.g.~\cite{Kazakov} and references therein.

In our model, it was possible to obtain exact results because the only non-trivial term entering expressions for the critical dimensions is $\gamma_{\nu}^*$ which is known exactly for both attractors~(\ref{IRGS2}) and~(\ref{IRGS3}).
The expressions for the critical dimensions in  Table~2 are universal in two respects: firstly, they are the same for any fixed point on the corresponding surface and secondly, they are determined solely by the spatial dimension $d$ for surface~(\ref{IRGS2}) and by $d$ and the exponent $y$ for~(\ref{IRGS3}).

Let us conclude with a brief discussion of a certain interesting ramification of our model. 

The NS equation with only the local term $k^2$ in the correlation function (\ref{noiseNS1}) has the form of a stochastic continuity equation. Then, the probability distribution of its equal-time correlations reduces to the well-known Maxwell's distribution with the action~$\sim{v^2}$; see, e.g.~\cite{FNS1,FNS} and Sec.~3.10 in~\cite{Red}.

The extended Pavlik's model acquires the form of a continuity equation in the absence of noise or, in its stochastic version, if the correlation function of the noise $\eta$ involves the additional factor $\partial^2 \sim k^2$ in comparison to (\ref{noise}); the case referred to as ``internal noise'' in the formulation of \cite{Diaz-Guilera}. 
Now, the full model can be interpreted as a critical dynamics for the equilibrium 
static model with the polynomial action functional $\sim \left(\sum_{n=2}h^n+v^2\right)$ with no derivatives (coefficients omitted).\footnote{The authors are indebted to M.Yu.~Nalimov for this observation. For discussion of critical dynamics, see, e.g. Chap.~5 in the monograph~\cite{Vasiliev}.} 
It is directly checked that such model is logarithmic in $d=0$, so that its critical 
behaviour for any $d>0$ is described by the corresponding free model with the critical dimensions equal to the canonical ones.

\bmhead{Acknowledgements}

This research was funded by the Russian Science Foundation grant number 24-22-00220, https://rscf.ru/en/project/24-22-00220/.

\section*{Declarations}
\begin{itemize}
\item Funding: This research was funded by the Russian Science Foundation grant number 24-22-00220, https://rscf.ru/en/project/24-22-00220/.
\item Competing interests: The authors have no competing interests to declare that are relevant to the content of this article.
\item Author contribution: All authors contributed equally to this work.
\item Ethics approval, Consent, Data, Material and Code availability: Not applicable.
\end{itemize}






\end{document}